\documentclass[sigconf]{aamas}

\usepackage{balance} 
\usepackage{algorithmic}
\usepackage{wrapfig}
\usepackage{booktabs}
\usepackage{graphicx}
\usepackage{subfigure}
\usepackage{makecell}
\usepackage{url}
\usepackage{amsmath}
\newtheorem{remark}{Remark}
\usepackage[most]{tcolorbox}
\usepackage{verbatim}
\usepackage{graphicx}
\usepackage{longtable}
\usepackage{multirow}
\usepackage{algorithm}

\doi{RQEQ9663}

\makeatletter
\gdef\@copyrightpermission{
  \begin{minipage}{0.2\columnwidth}
   \href{https://creativecommons.org/licenses/by/4.0/}{\includegraphics[width=0.90\textwidth]{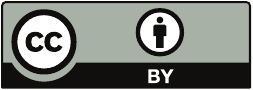}}
  \end{minipage}\hfill
  \begin{minipage}{0.8\columnwidth}
   \href{https://creativecommons.org/licenses/by/4.0/}{This work is licensed under a Creative Commons Attribution International 4.0 License.}
  \end{minipage}
  \vspace{5pt}
}
\makeatother

\setcopyright{ifaamas}
\acmConference[AAMAS '26]{Proc.\@ of the 25th International Conference
on Autonomous Agents and Multiagent Systems (AAMAS 2026)}{May 25 -- 29, 2026}
{Paphos, Cyprus}{C.~Amato, L.~Dennis, V.~Mascardi, J.~Thangarajah (eds.)}
\copyrightyear{2026}
\acmYear{2026}
\acmDOI{}
\acmPrice{}
\acmISBN{}

\acmSubmissionID{<<1531>>}

\title{MARLIN: Multi-Agent Reinforcement Learning with Murmuration Intelligence\\ and LLM Guidance for Reservoir Management}

\author{Heming Fu}
\affiliation{
  \institution{Department of Electrical and Computer Engineering}
  \city{Stony Brook University}
  \country{NY}
  }
\email{heming.fu@stonybrook.edu}

\author{Shan Lin}
\affiliation{
  \institution{Department of Electrical and Computer Engineering}
  \city{Stony Brook University}
  \country{NY}
  }
\email{shan.x.lin@stonybrook.edu}

\author{Guojun Xiong}
\affiliation{
  \institution{John Hopcroft Center}
  \city{School of Computer Science}
  \country{Shanghai Jiao Tong University}
  }
\email{gjxiong@sjtu.edu.cn}

\begin{abstract}\label{section:abstract}

Intensifying climate change and cascading uncertainties across interconnected reservoir networks pose escalating threats to global water security, demanding management systems that are both adaptive and scalable. Traditional centralized optimization becomes computationally intractable and brittle under real-world uncertainty, while existing reinforcement learning (RL) approaches are not designed for complex, multi-node hydrological systems. To address these challenges, we introduce \texttt{MARLIN}, a decentralized reservoir management framework that explicitly handles \textbf{dual-layer uncertainty}: (i) stochastic variability in physical water transfer and (ii) dynamic, human–environmental perturbations. 
\texttt{MARLIN} embeds bio-inspired \emph{alignment}, \emph{separation}, and \emph{cohesion} rules into a multi-agent RL (MARL) architecture to stabilize coordination under physical uncertainty. Additionally, external conditions such as weather forecasts, regulatory updates, and stakeholder preferences introduce \emph{unstructured textual information} that traditional models cannot process directly. To bridge this gap, we integrate a Large Language Model (LLM) that interprets such contextual information and dynamically adjusts the coordination parameters of the three murmuration rules, enabling rapid adaptation to evolving environmental and human requirements. 
Experiments on USGS data show that \texttt{MARLIN} improves uncertainty handling by 23\%, reduces computational cost by 35\%, and accelerates flood response by 68\%. The framework demonstrates excellent scalability, with emergent coordination patterns increasing super-linearly as the network expands while maintaining linear computational complexity. These results highlight \texttt{MARLIN}'s potential as a scalable and intelligent solution for adaptive water resource management and disaster prevention.
\end{abstract}


\newcommand{\BibTeX}{\rm B\kern-.05em{\sc i\kern-.025em b}\kern-.08em\TeX}

\begin{document}


\pagestyle{fancy}
\fancyhead{}


\maketitle 
\section{Introduction} \label{sec:abstract}

\begin{figure}[t]
\centering
\subfigure[Reservoir networks]{
  \includegraphics[width=0.4\linewidth]{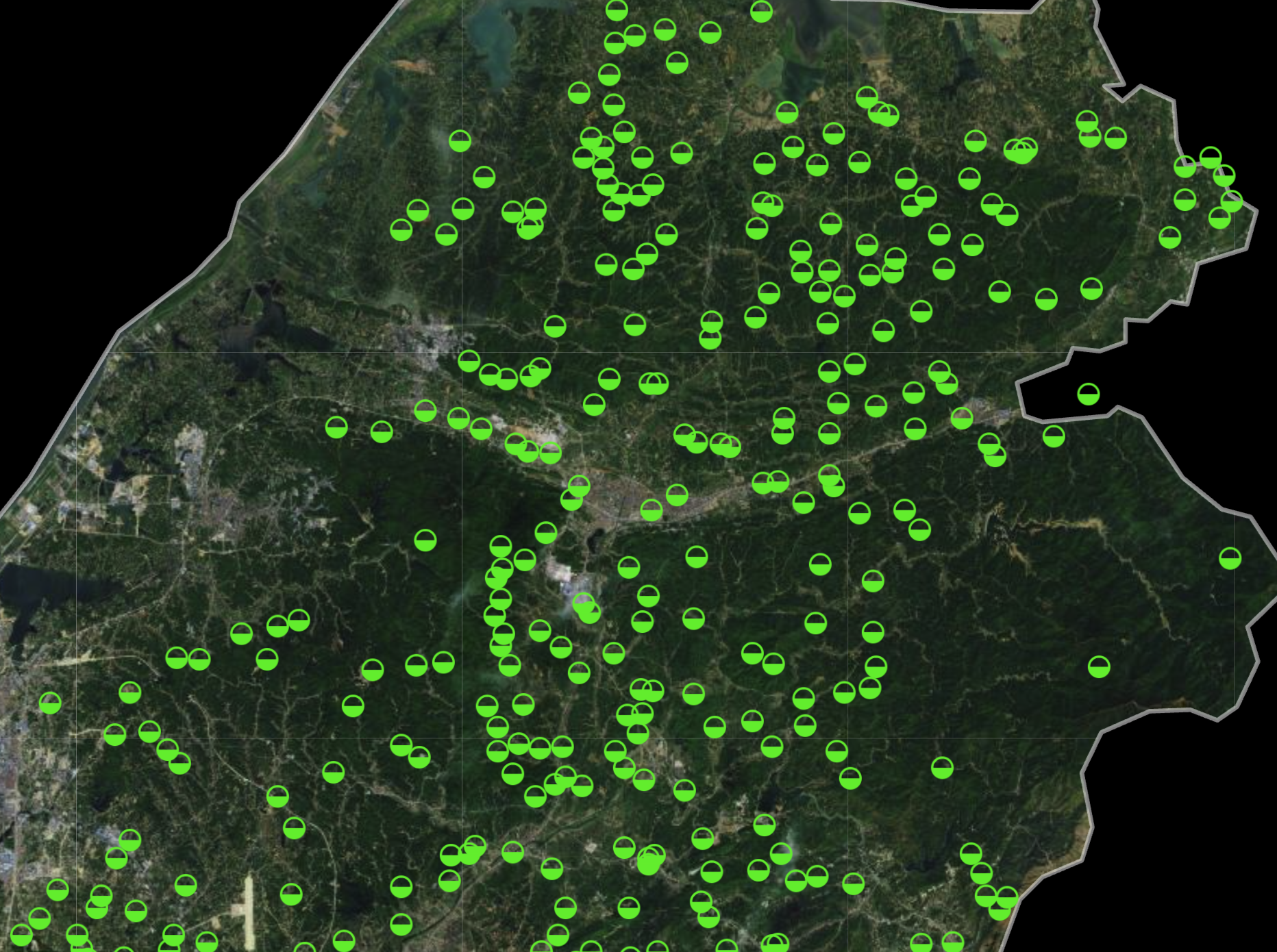}
  \label{fig:hunan_network}
  \vspace{-1em}
}
\subfigure[Local reservoir management
]{
  \includegraphics[width=0.53\linewidth]{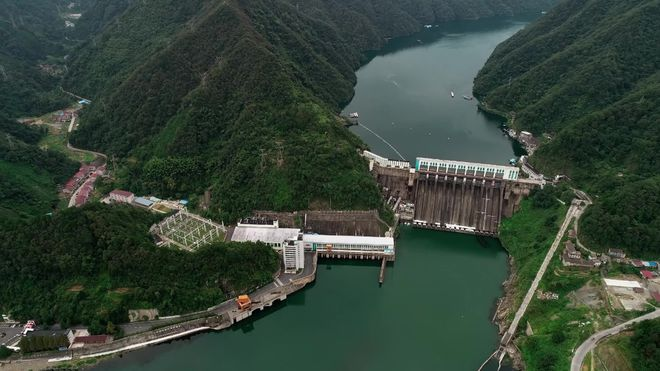}
  \label{fig:reservoir}
  \vspace{-1em}
}
\\
\subfigure[Global flood frequency map (1970-2018)]{
  \includegraphics[width=0.85\linewidth]{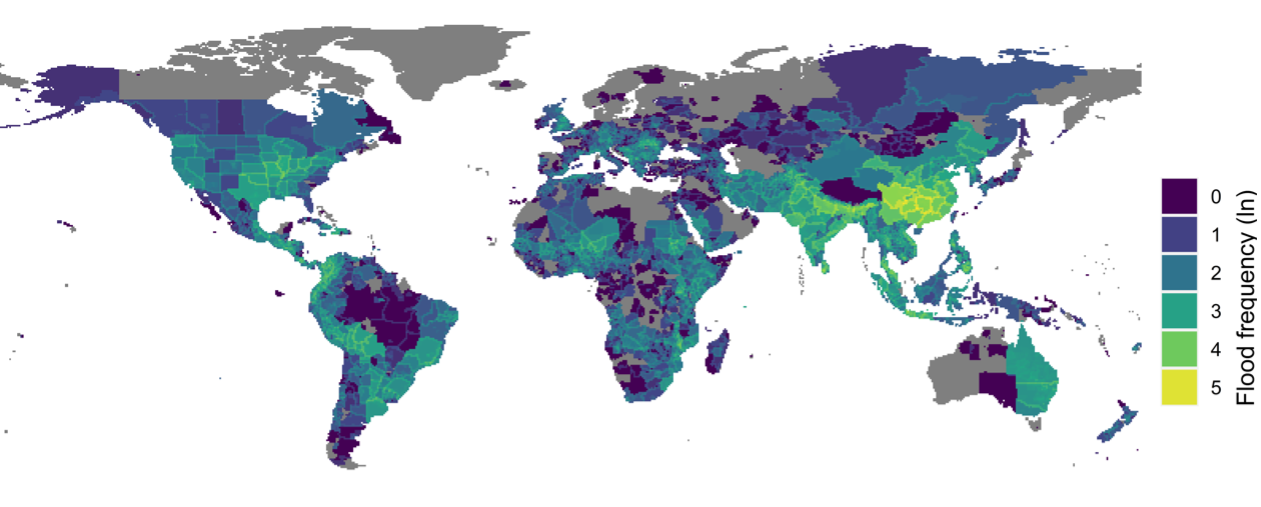}
  \label{fig:flood_map}
  \vspace{-1em}
}
\caption{Water resource management challenges: (a) complex networks of interconnected reservoirs requiring coordination, (b) individual reservoirs must be locally managed, and (c) high global flood frequency demanding  robust management systems. }
\label{fig:water_management}
\vspace{-1em}
\end{figure}

Water resource management is at a critical inflection point, as traditional engineering approaches are increasingly inadequate for the complexity and uncertainty of modern hydrological systems. Climate change has dramatically altered precipitation patterns (see Figure~\ref{fig:flood_map}), resulting in more frequent and severe extreme weather events that disproportionately impact vulnerable populations. The consequences are global: between 2000 and 2019, floods affected over 1.65 billion people and caused an estimated 651 billion in economic losses~\cite{yu2022extreme}, while droughts impacted 1.43 billion people and resulted in damages exceeding 130 billion~\cite{corbari2024multi}. Events that were once considered century-scale extremes now occur every 20 years in many regions, and flash flood frequency has surged by 21\% over the past decade~\cite{axios2025flashflood}.
Recent disasters underscore the urgent need for adaptive water management systems. The 2021 European floods resulted in over 200 deaths and approximately $€$50 billion in damages \cite{collaboration2023coordination}, and New York's 2023 extreme rainfall events both revealed critical failures in infrastructure and cross-jurisdictional coordination~\cite{lander2024readyforrain}. Similarly, Texas winter storms repeatedly threaten water security by disabling power-dependent systems~\cite{fechter2021texasstorm,sugg2023cascading}, while a historic megadrought has pushed Lakes Mead and Powell below 30\% capacity, jeopardizing water supplies for 40 million people~\cite{cbs2021megadrought,calmatters2024cuts}. These are not isolated incidents but symptoms of systemic vulnerabilities. 
 A systematic study of US dam failures, including the 2020 Michigan breach and the 2017 Oroville spillway collapse, concluded that current hydrologic design and operation protocols are often inadequate for managing the clustered, extreme weather events of a changing climate~\cite{Hwang2024DamFailure}.

\begin{figure}[t]
    \centering
    \includegraphics[width=1\linewidth]{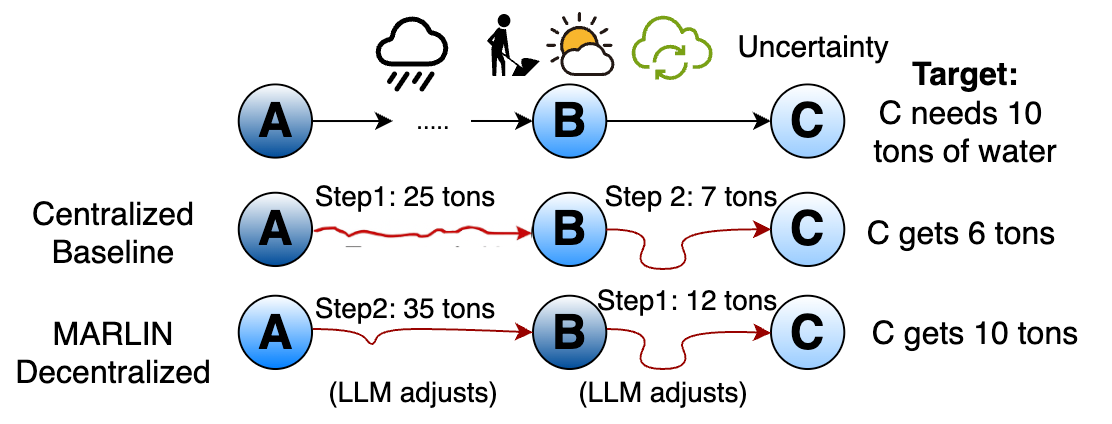}
    \caption{
    \textbf{Illustration of centralized vs. decentralized coordination under uncertainty.}
    The target is for node~C to receive 10~tons of water. 
    In the \textit{centralized baseline}, node~A releases 25~tons based on static planning, 
    yet due to environmental variations like rainfall, 
    node~C ultimately receives only 6~tons. 
    In contrast, \textit{MARLIN decentralized coordination} allows each node to adapt through local feedback: 
    node~B first releases 12~tons, and node~A subsequently compensates by adjusting its output to 35~tons, 
    guided by LLM-assisted local adjustments. 
    This closed-loop adaptation enables node~C to obtain 10~tons despite environmental uncertainty.
    }
    \label{fig:marlin_baseline_comparison}
\end{figure}

These widespread failures stem from two fundamental limitations of current centralized management paradigms.
First, as reservoir networks grow, computational complexity scales as $O(N^3)$, making real-time response to rapidly changing conditions impractical~\cite{koo2025flood,lin2020multi,wang2020mpc}. 
Second, these centralized models are highly vulnerable to cascading uncertainty from dual sources: (i) \emph{physical water transfer uncertainty} arising from evaporation, seepage, and variable channel conditions during water transport; and (ii) \emph{environmental uncertainty} encompassing seasonal variations (e.g., spring agricultural irrigation demanding 60\% of annual water allocation), extreme weather events, and human activities ranging from routine operations to emergency situations (e.g., industrial accidents requiring immediate water supply for cooling, as seen in the 2023 Fukushima release~\cite{wikipedia_fukushima_discharge}). These diverse environmental factors create complex, time-varying demands that traditional models cannot anticipate. 
For instance, minor errors compound catastrophically through network propagation: assuming independent 7\% allocation uncertainty at each node, the cumulative error grows as $\sigma_{\text{cumulative}} = \sigma_{\text{base}}\sqrt{n}$, reaching 22\% after just 10 nodes and exceeding 40\% in networks with 30+ interconnected reservoirs, which are typical of major river basins. Under correlated environmental conditions, this amplification accelerates further, rendering centralized predictions unreliable, as illustrated in Fig~\ref{fig:marlin_baseline_comparison}.
Additionally, multi-agent reinforcement learning (MARL) provides a distributed alternative but remains limitations to uncertainty~\cite{zhang2020robust,he2024robust,rashid2018qmix, yu2021surprising}, which often suffer from training instability and poor convergence under cascading uncertainties, resulting in unsafe, oscillatory behaviors in practice~\cite{he2023robust}.

\begin{wrapfigure}{r}{0.20\textwidth} 
  \centering
\includegraphics[width=0.20\textwidth]{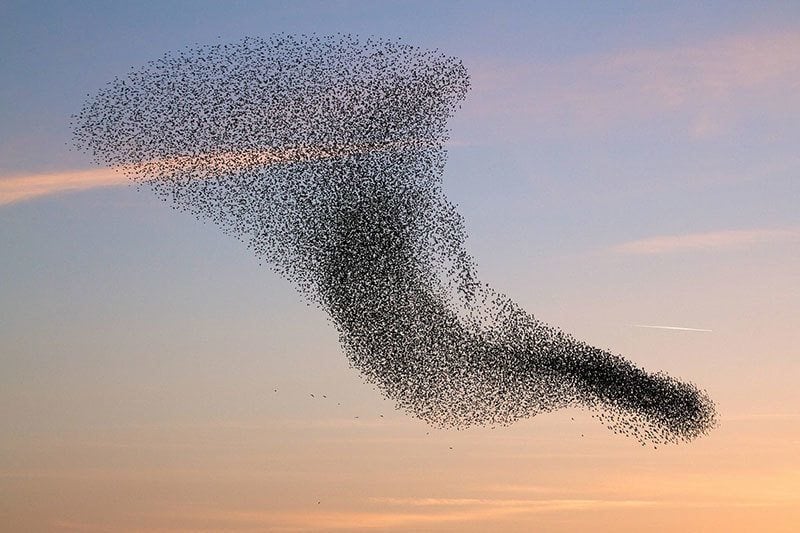}
  \caption{Starling Murmuration: emergent intelligence from simple local coordination rules.}
  \label{fig:murmuration}
  \vspace{-0.4cm}
\end{wrapfigure}

Motivated by nature's ability to achieve robust coordination under uncertainty, we draw inspiration from starling murmurations (Figure~\ref{fig:murmuration}). These flocks show how complex, adaptive organization can emerge from simple, local rules—\emph{\textbf{alignment}}, \emph{\textbf{separation}}, and \emph{\textbf{cohesion}}—without any centralized control~\cite{reynolds1987flocks,hildenbrandt2010self,cavagna2010scale}. 
We translate this principle of emergent intelligence into \texttt{MARLIN}, a decentralized framework that integrates these bio-inspired rules with reinforcement learning in two ways. First, to manage physical water transfer uncertainty, \texttt{MARLIN} maps these bio-inspired rules to reservoir control actions and directly incorporates their signals into the policy gradients of a specialized MARL architecture. 
\begin{figure}[t]
  \centering
  \includegraphics[width=1.05\linewidth]{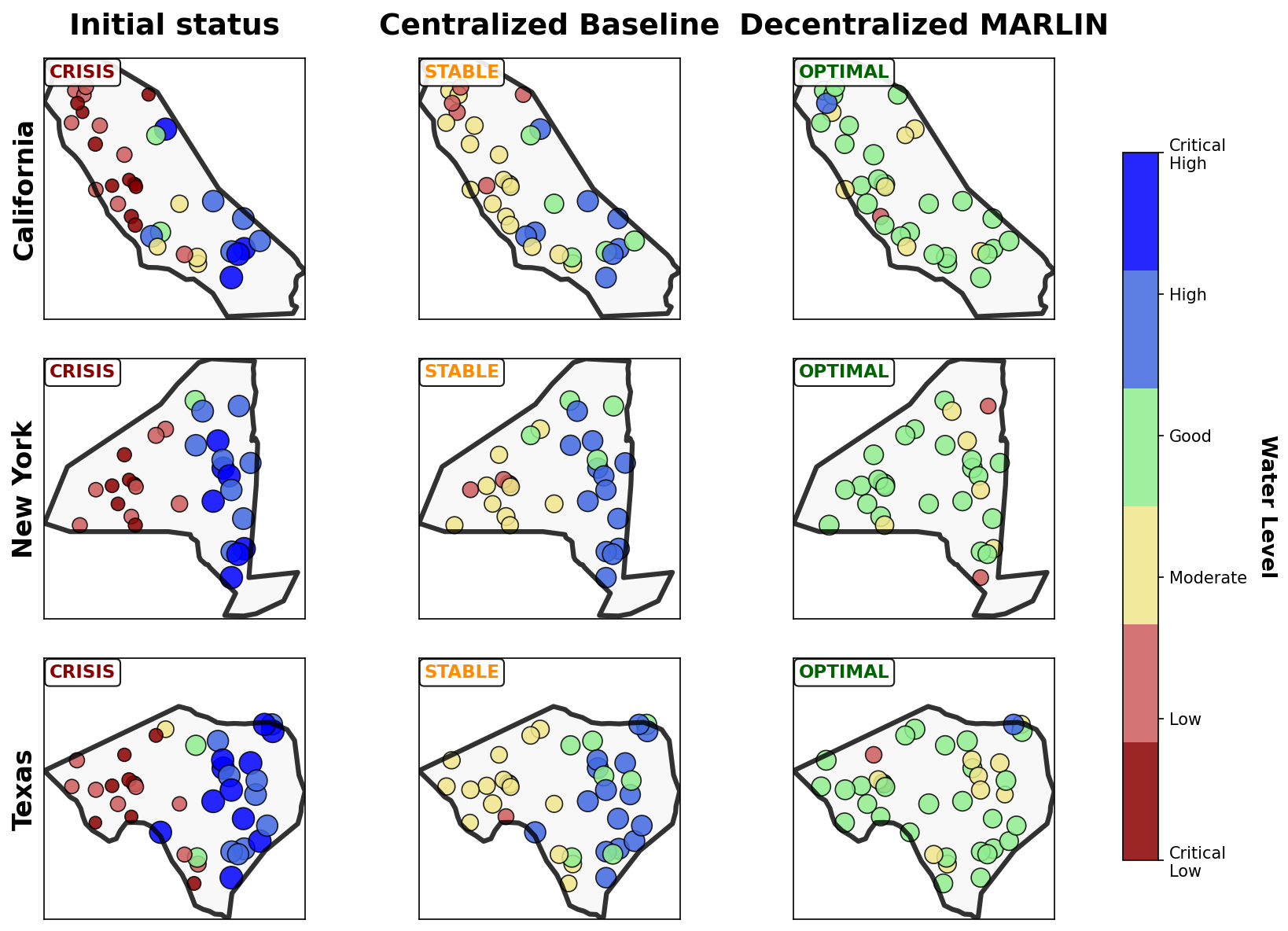}
  \caption{
  \textbf{Statewide reservoir status under uncertainty: Initial vs.\ Centralized Baseline vs.\ Decentralized \texttt{MARLIN}.}
  Each panel shows reservoir water levels.
  The centralized baseline yields only modest stabilization, whereas \texttt{MARLIN} achieves near-optimal balance with lower variance and better regional coordination.
  }
  \label{fig:intro_marlin_grid}
  \vspace{-0.8em}
\end{figure}
This novel integration promotes stable learning while fostering emergent, system-wide safety and responsiveness. Second, to address the more complex environmental uncertainties, \texttt{MARLIN} integrates a Large Language Model (LLM). This allows the system to process diverse, unstructured textual information from weather forecasts to regulatory documents and translate its knowledge into adaptive reward-shaping signals, guiding agents in dynamic contexts that are intractable for traditional optimization. 
As shown in Fig.~\ref{fig:intro_marlin_grid}, reservoir states progress from crisis (Initial) to modest stability under a centralized baseline, and to near-optimal balance with \texttt{MARLIN}.


In summary, our main contributions are:

\begin{itemize}
\item[$\triangleright$] We propose a novel bio-inspired MARL system, \texttt{MARLIN}, that applies starling murmuration principles to achieve emergent and scalable coordination from simple local interactions in decentralized water management.

\item[$\triangleright$] We develop a dual-layer uncertainty handling mechanism, where simple local coordination rules manage physical water transfer uncertainty, while LLM-guided reward shaping adapts to complex environmental fluctuations and human preferences.

\item[$\triangleright$] Experiments on USGS watershed data demonstrate that \texttt{MARLIN} achieves substantial improvements over baselines: 23\% better uncertainty handling, 35\% lower computation cost, 68\% faster flood response, and 42\% improvement in regional water balance, underscoring its potential for safeguarding vulnerable communities against water-related disasters.
\end{itemize}

\section{Related Work} \label{sec:related}

\noindent\textbf{Water Management and Multi-Agent Systems.}
Classical water management relied on deterministic optimization like dynamic programming~\cite{yakowitz1982dp,li2014parallel} and network flow models~\cite{tahiri2022network}, suffering from exponential complexity and static assumptions. MPC became dominant in operational systems but demonstrates critical limitations during rapid changes. The 2011 Mississippi floods exemplified delayed recalibration contributing to \$2.8 billion avoidable damage~\cite{mpr2013flood}.
MARL emerged as a distributed alternative. MADDPG~\cite{lowe2017multi} uses centralized critics with decentralized actors, QMIX~\cite{rashid2018qmix} employs value decomposition with monotonic mixing, and MAPPO~\cite{yu2021surprising} adapts PPO to multi-agent settings. However, these methods face training instability under non-stationary environments and assume coordination emerges naturally from individual optimization, overlooking explicit coordination needed for uncertainty handling~\cite{xiong2025MurmuRL}.

\noindent\textbf{Bio-Inspired Coordination.}
Reynolds formalized flocking behavior into alignment, separation, and cohesion rules~\cite{reynolds1987flocks}. Theoretical analysis revealed phase transitions between ordered and disordered states~\cite{chate2014insect,vicsek1995novel}, while empirical studies confirmed scale-free correlations and rapid information propagation~\cite{cavagna2010scale}. These inspired algorithms like Particle Swarm Optimization~\cite{kennedy1995particle} and practical swarm robotics applications~\cite{rubenstein2014programmable,Poli2007PSOAnalysis}. However, most focus on continuous motion rather than discrete decision-making required in infrastructure control.

\noindent\textbf{LLMs for Adaptive Control.}
Recent work explores LLMs for reward design and adaptation~\cite{xiong2025rulebottleneck, xiong2025vortexaligningtaskutility}. Eureka~\cite{ma2023eureka} uses GPT-4 for automated reward generation, while L2R~\cite{goyal2019using,goyal2021pixl2r} enables natural language reward specification. However, these approaches primarily focus on single-agent settings or assume full observability, limiting their applicability in decentralized multi-agent systems.
\section{Problem Formulation}\label{sec:problem}

\begin{figure}[h]
\centering
\includegraphics[width=0.4\textwidth]{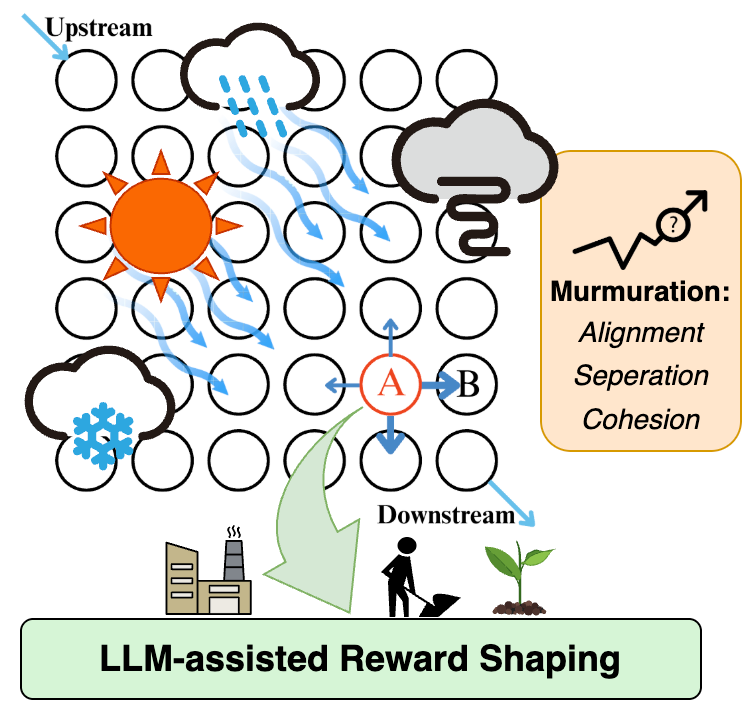}
\caption{
Illustration of the multi-layer reservoir network system under cascading uncertainty. 
Upstream environmental disturbances (e.g., heat, rainfall, freezing, human activities) dynamically affect inflows, leading to transfer uncertainty among interconnected reservoirs. 
Each node (e.g., node A) operates under both environmental and flow-level uncertainties. 
The right panel shows the \textit{murmuration-inspired} coordination rules (alignment, separation, and cohesion) that address the \emph{Level 1: Physical Transfer Uncertainty.} 
The LLM-assisted reward shaping layer translates human, industrial, and ecological objectives into adaptive decision feedback, addressing \emph{Level 2: Environmental and Human Modulation.}
}
\label{fig:network_model}
\end{figure}

We model modern reservoir networks as distributed decision systems operating under dual-source uncertainty—physical transfer variability and environmental-human factors—that propagates through interconnected hydrological channels, as shown in Fig~\ref{fig:network_model}.

\paragraph{\textbf{Network Model and Stochastic Dynamics}}
We model the reservoir network as a directed graph $\mathcal{G} = (\mathcal{V}, \mathcal{E})$, where $\mathcal{V} = \{v_1, v_2, \ldots, v_N\}$ represents $N$ reservoir nodes and $\mathcal{E} \subseteq \mathcal{V} \times \mathcal{V}$ denotes water flow channels. Each reservoir $i$ maintains state vector $\mathbf{s}_i(t) = [h_i(t), q_{\text{in},i}(t), q_{\text{out},i}(t), \boldsymbol{\omega}_i(t), d_i(t)]^\intercal$. Here, $h_i(t)$ represents water level in meters determining storage capacity and downstream pressure (for instance, Lake Mead's water level dropping below 320 meters triggers mandatory conservation measures affecting 40 million people), $q_{\text{in/out},i}(t)$ aggregate all incoming and outgoing flows including controlled releases (typically 200-500 m³/s during normal operations) and spillway overflow (can exceed 10,000 m³/s during extreme floods as seen in the 2017 Oroville crisis~\cite{ift2018oroville}), $\boldsymbol{\omega}_i(t)$ captures local weather conditions (temperature affecting evaporation, precipitation intensity, humidity) collected from NOAA stations, and $d_i(t)$ represents time-varying demand from agricultural irrigation (peaks at 800 m³/s during California's summer growing season), urban consumption, and industrial processes.

The water level dynamics follow:
\begin{align}
\frac{dh_i(t)}{dt} = \frac{1}{A_i}\left[\sum_{j \in \mathcal{N}_i^{\text{up}}} f_{ji}(t) \!-\!\!\!\! \sum_{k \in \mathcal{N}_i^{\text{down}}} f_{ik}(t) + q_{\text{ext},i}(t) + \eta_i(t)\right],
\end{align}
where $A_i$ is reservoir surface area, $\mathcal{N}_i^{\text{up/down}}$ denote upstream/downstream neighbors, $f_{ij}(t)$ represents actual water transfer between reservoirs (defined below), $q_{\text{ext},i}(t)$ captures external inflows from rainfall and tributaries, and $\eta_i(t) \sim \mathcal{N}(0, \sigma_{\eta}^2)$ models measurement noise. The control actions relate to flows as $q_{\text{out},i}(t) = \sum_{k \in \mathcal{N}_i^{\text{down}}} a_{i \rightarrow k}(t)$ where $a_{i \rightarrow k}(t)$ is the controlled release from reservoir $i$ to $k$, and $q_{\text{in},i}(t) = \sum_{j \in \mathcal{N}_i^{\text{up}}} f_{ji}(t)$ represents actual received water after transfer losses.

\noindent\paragraph{\textbf{Dual-Source Uncertainty}}
The critical challenge lies in cascading uncertainty from two distinct sources.

\noindent\emph{Level 1: Physical Transfer Uncertainty.} When reservoir $i$ releases volume $a_{i \rightarrow j}(t)$ toward downstream reservoir $j$, the actual water transfer follows:
\begin{align}
f_{ij}(t) = \alpha_{ij}(t) \cdot a_{i \rightarrow j}(t) + \epsilon_{ij}(t),
\end{align}
where $\alpha_{ij}(t) \in [\epsilon, 1]$ represents time-varying channel efficiency accounting for evaporation, seepage, and channel conditions, with $\epsilon = 0.1$ ensuring minimum 10\% efficiency even in extreme conditions, while $\epsilon_{ij}(t) \sim \mathcal{N}(0, \sigma_{\text{base}}^2)$ captures random fluctuations. During normal conditions, concrete-lined channels maintain $\alpha_{ij} \approx 0.95$, but natural earthen channels drop to $\alpha_{ij} \approx 0.70$ due to seepage through permeable soils. Extreme heat waves can reduce efficiency by an additional 15-20\% through evaporation.

\noindent\emph{Level 2: Environmental and Human Modulation.} The channel efficiency is affected by environmental and human factors:
$$
\begin{aligned}
\alpha_{ij}(t) \!=\! 
\min\!\Big(&1,\ 
\max\!\big(\epsilon,\ 
\alpha_{\text{nominal}} \!\cdot\! 
\big(1 
-\ \gamma_{\text{env}}(\boldsymbol{\omega}_i(t), \boldsymbol{\omega}_j(t)) \\
&-\ \gamma_{\text{human}}(d_i(t), d_j(t))\big)
\big)\Big),
\end{aligned}
$$
where $\gamma_{\text{env}}$ captures weather-induced losses (higher evaporation during heatwaves when temperatures in $\boldsymbol{\omega}_i(t)$ exceed 40°C can increase $\gamma_{\text{env}}$ by 0.3), and $\gamma_{\text{human}}$ captures both predictable patterns from demand $d_i(t)$ and unpredictable interventions from textual sources $\mathcal{T}(t)$. These textual sources include news articles ("Fukushima plant~\cite{wikipedia:fukushima} requires 500,000 gallons per minute for emergency cooling"), regulatory bulletins ("Stage 3 drought restrictions effective immediately"), and social media alerts ("Chemical spill reported upstream of intake").

\noindent\paragraph{\textbf{Uncertainty Propagation}}
For a cascade of $n$ reservoirs, cumulative uncertainty compounds as:
\begin{align}
\text{Var}[h_n(t)] = \sum_{i=1}^{n-1} \left(\prod_{j=i+1}^{n} \alpha_{j,j-1}^2(t)\right) \sigma_{\text{base}}^2 + \sigma_{\eta}^2,
\end{align}
showing how upstream uncertainties are amplified by the product of channel efficiencies. When $\alpha_{ij}(t) \approx 0.9$ (typical conditions), uncertainty grows moderately, but during extreme events with $\alpha_{ij}(t) \approx 0.5$, downstream uncertainty explodes exponentially. In the Colorado River's 15-reservoir cascade, even with individual $\alpha_{ij} = 0.93$, the compound efficiency to Lake Mead drops to 0.35, meaning 65\% uncertainty in delivery.

\noindent\paragraph{\textbf{Multi-Objective Control Problem}}
We formulate reservoir management as a multi-objective optimization problem explicitly designed to balance four interacting goals—\emph{safety}, \emph{supply}, \emph{ecology}, and \emph{efficiency}—that together define the operational priorities of the system, i.e.,
\begin{align}
J_{\text{safety}}(\pi) &= -\mathbb{E}_{\pi}\left[\sum_{t=0}^{T} \sum_{i=1}^{N} \lambda_{\text{flood},i} \cdot \mathbb{I}[h_i(t) > h_{\text{safe},i}]\right], \\
J_{\text{supply}}(\pi) &= \mathbb{E}_{\pi}\left[\sum_{t=0}^{T} \sum_{i=1}^{N} \min\left(1, \frac{q_{\text{out},i}(t)}{d_i(t)}\right)\right], \\
J_{\text{ecology}}(\pi) &= \mathbb{E}_{\pi}\left[\sum_{t=0}^{T} \min\left(1, \frac{\sum_{i \in \mathcal{V}} q_{\text{out},i}(t)}{F_{\text{eco}}(t)}\right)\right], \\
J_{\text{efficiency}}(\pi) &= -\mathbb{E}_{\pi}\left[\sum_{t=0}^{T} \sum_{i=1}^{N} c_{\text{op},i} \cdot q_{\text{out},i}(t)\right],
\end{align}
where $\lambda_{\text{flood},i}$ weights flood risk priority for populated areas, $h_{\text{safe},i}$ is the safety threshold for reservoir $i$, $F_{\text{eco}}(t)$ specifies minimum ecological flow requirements, and $c_{\text{op},i}$ represents operational cost per unit water released. In addition,
each reservoir $i$ observes only local information:
\begin{equation}
\mathcal{O}_i(t) = \{\mathbf{s}_i(t), \{\mathbf{s}_j(t-\tau)\}_{j \in \mathcal{N}_i^{\text{up}} \cup \mathcal{N}_i^{\text{down}}}, \boldsymbol{\omega}_{\text{forecast},i}(t{:}t+H)\},
\end{equation}
where $\tau$ is communication delay and $\boldsymbol{\omega}_{\text{forecast},i}(t{:}t+H)$ provides $H$-hour weather forecasts.
The optimization seeks decentralized policies $\pi_i: \mathcal{O}_i \rightarrow \{a_{i \rightarrow k}(t)\}_{k \in \mathcal{N}_i^{\text{down}}}$ maximizing:
\begin{align}
\max_{\pi} \quad & \mathbf{J}(\pi) = [J_{\text{safety}}, J_{\text{supply}}, J_{\text{ecology}}, J_{\text{efficiency}}]^\intercal \\
\text{s.t.} \quad & 0 \leq a_{i \rightarrow k}(t) \leq a_{\text{max},i}, \quad h_{\min} \leq h_i(t) \leq h_{\max}.
\end{align}

\begin{remark}
This formulation reveals three fundamental challenges: (1) Cascading uncertainty amplification where errors grow exponentially through the network under reduced channel efficiency, making centralized prediction unreliable; (2) Multi-objective coupling where flood safety conflicts with supply reliability, especially during extreme events; (3) Information asymmetry where environmental factors arrive as structured data ($\boldsymbol{\omega}_i(t)$) while human interventions manifest as demand spikes ($d_i(t)$) or unstructured text requiring different processing paradigms. These challenges motivate \texttt{MARLIN}'s dual approach in Section~\ref{sec:methodology}—bio-inspired coordination to handle physical uncertainty through local rules and LLM-guided adaptation to process contextual information about environmental and human factors.
\end{remark}

\section{Methodology}\label{sec:methodology}

\begin{figure*}[t]
\centering
\includegraphics[width=1\linewidth]{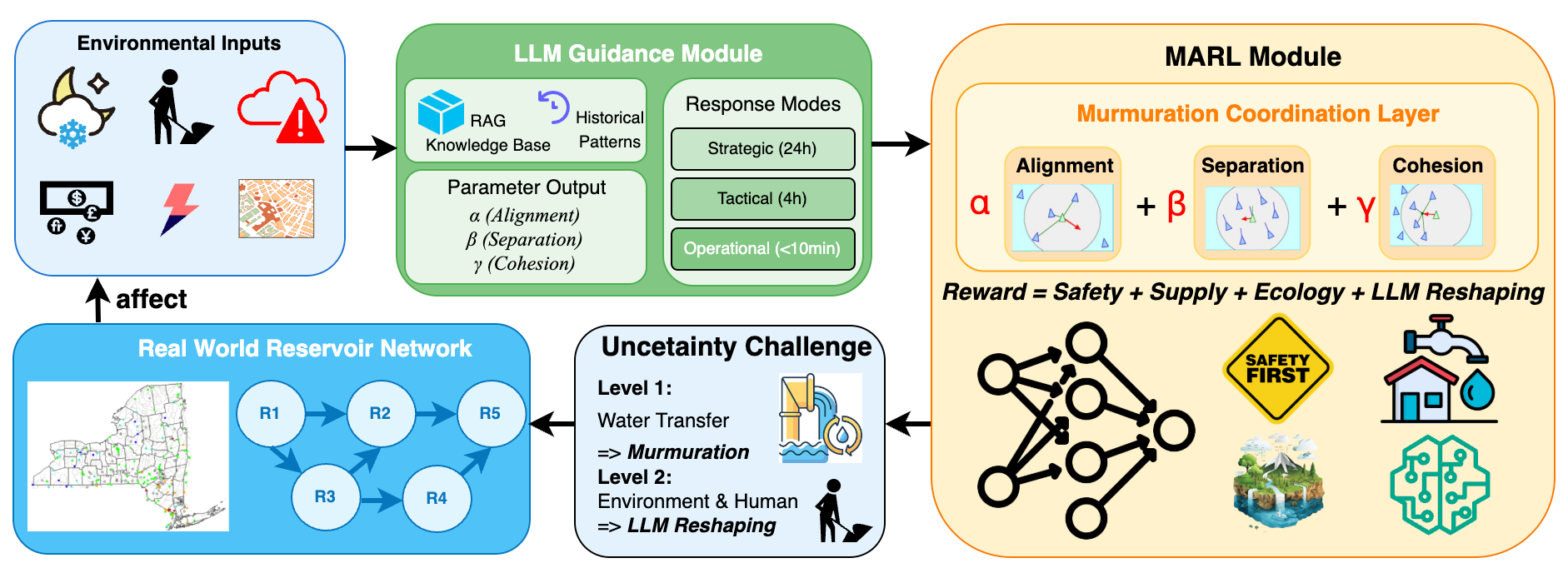}
\caption{
\textbf{MARLIN system overview.} 
The pipeline proceeds left-to-right: \textit{Environmental Inputs} (weather, human activities, emergencies, etc.) feed the \textit{LLM Guidance Module}, which parses context (RAG knowledge base, historical patterns) and outputs response modes (strategic/tactical/operational) and murmuration parameters ($\alpha, \beta, \gamma$). 
The \textit{MARL Module} integrates a \textit{Murmuration Coordination Layer} (Alignment, Separation, Cohesion) and optimizes a reward composed of \textit{Safety + Supply + Ecology + LLM Reshaping}. 
The \textit{Real-World Reservoir Network} executes decentralized actions and provides feedback, while the \textit{Uncertainty Challenge} highlights two layers: (L1) water-transfer variability handled by murmuration rules and (L2) environment/human factors handled by LLM-based reshaping.
}
\label{fig:marlin-overview}
\end{figure*}

\texttt{MARLIN} addresses the dual-source uncertainty through a hierarchical approach: bio-inspired coordination rules handle Level 1 physical transfer uncertainty through local mathematical operations, while LLM-guided adaptation processes Level 2 environmental-human uncertainty from textual sources $\mathcal{T}(t)$, as shown in Fig~\ref{fig:marlin-overview}.

\subsection{Murmuration-Inspired Coordination Layer}

To address the cascading physical uncertainty where variance grows as $\text{Var}[h_n(t)] = \sum_{i=1}^{n-1} \left(\prod_{j=i+1}^{n} \alpha_{j,j-1}^2(t)\right) \sigma_{\text{base}}^2$, we adapt three starling murmuration rules that create robust local consensus without requiring global information.

\noindent\paragraph{\textbf{Adaptive Alignment Rule}} 
This rule encourages coordinated release actions among neighboring reservoirs to maintain consistent flow despite uncertainty:
\begin{align}
    \mathcal{L}_{\text{align},i}(t)\! =\!\! \sum_{j \in \mathcal{N}_i^{\text{up}} \cup \mathcal{N}_i^{\text{down}}}\!\!\! w_{ij}(t) \cdot \left\|\sum_{k \in \mathcal{N}_i^{\text{down}}}\!\!\! a_{i \rightarrow k}(t) - \bar{a}_{ij}(t)\right\|^2,
\end{align}
where $\bar{a}_{ij}(t)$ is the average release between reservoirs accounting for communication delay $\tau$, and weights are computed as:
\begin{align}
w_{ij}(t) = \frac{\exp(-\beta_d \delta_{ij} - \beta_e \|\boldsymbol{\omega}_i(t) - \boldsymbol{\omega}_j(t)\|_2)}{\sum_{k \in \mathcal{N}_i^{\text{up}} \cup \mathcal{N}_i^{\text{down}}} \exp(-\beta_d \delta_{ik} \!-\! \beta_e \|\boldsymbol{\omega}_i(t) \!-\! \boldsymbol{\omega}_k(t)\|_2)},
\end{align}
where $\delta_{ij}$ is geographic distance between reservoirs and $\boldsymbol{\omega}_i(t)$ is the weather vector from state $\mathbf{s}_i(t)$. This weighting ensures reservoirs experiencing similar weather conditions coordinate more strongly, directly addressing correlated uncertainty during regional events.

\noindent\paragraph{\textbf{Strategic Separation Rule}}
To prevent catastrophic cascading failures when channel efficiency $\alpha_{ij}(t)$ drops below 0.5, this rule maintains action diversity:
\begin{align}
    \mathcal{L}_{\text{sep},i}(t)\!\! =\!\!\! \sum_{j \in \mathcal{N}_i^{\text{up}} \cup \mathcal{N}_i^{\text{down}}}\!\!\!\! \phi_{\text{sep}}\left(\left\|\sum_{k} a_{i \rightarrow k}(t) \!-\!\! \sum_{k} a_{j \rightarrow k}(t-\tau)\right\|_2; \rho_i(t)\right),
\end{align}
where $\phi_{\text{sep}}(x; \rho) = \exp(-x^2/\rho^2)$ penalizes similar total releases, and the diversity radius adapts to local uncertainty:
\begin{align}
\rho_i(t) = \rho_{\text{base}} \cdot \left(1 + \text{CV}[\{h_j(t-\tau)\}_{j \in \mathcal{N}_i^{\text{up}} \cup \mathcal{N}_i^{\text{down}}}]\right),
\end{align}
with CV denoting coefficient of variation. When water levels show high variability, $\rho_i(t)$ increases, encouraging more diverse strategies.

\noindent\paragraph{\textbf{Ecological Cohesion Rule}}
This rule ensures collective releases meet ecological requirements $F_{\text{eco}}(t)$ from the objective $J_{\text{ecology}}$:
\begin{align}
    \mathcal{L}_{\text{coh},i}(t) = \lambda_{\text{eco},i}(t) \cdot \left\|\sum_{j \in \mathcal{P}_i} q_{\text{out},j}(t-\tau) - \frac{F_{\text{eco}}(t)}{|\mathcal{P}_i|}\right\|^2,
\end{align}
where $\mathcal{P}_i \subseteq \mathcal{V}$ represents reservoirs in the same ecological region.

\subsection{MARL with Coordination Integration}
Training follows the Centralized Training with Decentralized Execution (CTDE) paradigm~\cite{CTDE}, where global information is used only during training, and each reservoir executes its policy using strictly local observations at deployment.Building upon the coordination rules, each agent $i$ constructs its policy using the local observation $\mathcal{O}_i(t)$ and an enhanced state representation:
\begin{equation}
\mathbf{s}_i^{\text{MARL}}(t) = 
\left[
\begin{aligned}
&\mathbf{s}_i(t),\
\text{GNN}_{\theta}\!\big(\{\mathbf{s}_j(t{-}\tau),\, \mathbf{e}_{ij}\}_{j \in \mathcal{N}_i}\big),\\
&\text{LSTM}_{\phi}\!\big(\{\mathbf{s}_i(k)\}_{k=t-K}^{t}\big),\
\boldsymbol{\omega}_{\text{forecast},i}(t{:}t{+}H)
\end{aligned}
\right]^{\!\top},
\label{eq:state-marl}
\end{equation}
where the GNN module encodes delayed neighbor information with edge features 
$\mathbf{e}_{ij} = [\hat{\alpha}_{ij}(t), \delta_{ij}]$ representing the estimated channel efficiency and flow delay.  
The LSTM captures temporal correlations from the past $K$ steps, while $\boldsymbol{\omega}_{\text{forecast},i}(t{:}t{+}H)$ provides short-term weather forecasts to anticipate upcoming inflow variations.

\paragraph{Murmuration Gradient Integration.}
The policy network incorporates coordination feedback from the murmuration layer through gradient-based modulation:
\begin{equation}
\mathbf{h}_i^{(2)} 
= \mathbf{h}_i^{(1)} 
+ \xi \cdot \text{MLP}\!\big([\nabla_{a_i}\mathcal{L}_{\text{align},i},\ 
\nabla_{a_i}\mathcal{L}_{\text{sep},i},\ 
\nabla_{a_i}\mathcal{L}_{\text{coh},i}]^{\top}\big),
\label{eq:murm-grad}
\end{equation}
where $\mathbf{h}_i^{(1)}$ and $\mathbf{h}_i^{(2)}$ denote consecutive hidden representations of the policy network, and $\xi$ regulates the strength of bio-inspired influence.  
This integration allows directional guidance from the murmuration rules: alignment promotes synchronization with neighbors, separation enforces diversity to avoid collective failures, and cohesion drives consistency with ecological objectives.

\paragraph{Training Objective.}
Training follows a modified PPO formulation that jointly optimizes performance and coordination:
\begin{equation}
\begin{aligned}
J_i^{\text{PPO}} =\ 
&\mathbb{E}_{\mathcal{B}_i}\!\left[
\min\!\big(
r_t(\theta_i)A_t,\ 
\text{clip}(r_t(\theta_i), 1{-}\epsilon, 1{+}\epsilon)A_t
\big)
\right] \\
&\quad -\ 
\beta_{\text{mur}}\,
\mathcal{L}_{\text{total},i}(t),
\end{aligned}
\label{eq:ppo-murm}
\end{equation}
where $r_t(\theta_i)=\frac{\pi_{\theta_i}(a_t|s_t)}{\pi_{\theta_i}^{\text{old}}(a_t|s_t)}$ is the probability ratio, $A_t$ denotes the advantage, and $\mathcal{B}_i$ is the experience batch.  
The murmuration penalty aggregates all coordination objectives:
\begin{equation}
\mathcal{L}_{\text{total},i}(t)
= \kappa_{\text{align}}\,\mathcal{L}_{\text{align},i}
+ \kappa_{\text{sep}}\,\mathcal{L}_{\text{sep},i}
+ \kappa_{\text{coh}}\,\mathcal{L}_{\text{coh},i},
\end{equation}
with $\kappa_{\text{align}} + \kappa_{\text{sep}} + \kappa_{\text{coh}} = 1$.
Here, $\beta_{\text{mur}}$ adaptively balances performance optimization and coordination strength—rising during emergencies to emphasize collective safety, and decreasing during stable periods to favor local efficiency.

\subsection{LLM-Guided Adaptive Reward Shaping}

To address Level~2 environmental--human uncertainty arising from textual sources $\mathcal{T}(t)$, an LLM module is employed to transform unstructured information into structured parameter adjustments for the reinforcement learning agents.

\paragraph{Context-Integrated Reward Function.}
The instantaneous reward incorporates both standard operational components and LLM-informed contextual shaping:
\begin{equation}
\begin{aligned}
R_i(t) =\ 
&r_{\text{safety},i}\!\big(h_i(t)\big)
+ r_{\text{supply},i}\!\big(q^{\text{out}}_i(t), d_i(t)\big) \\
&+ r_{\text{eco},i}\!\big(q^{\text{out}}_i(t)\big)
+ R_{\text{shaped},i}\!\big(\boldsymbol{\psi}(t)\big),
\end{aligned}
\label{eq:reward}
\end{equation}
where each $r_k$ corresponds to a primary objective component from $J_k$, and $R_{\text{shaped},i}(\boldsymbol{\psi}(t))$ provides adaptive adjustments derived from LLM-processed context.

\paragraph{Text-to-Parameter Translation.}
The LLM parses incoming textual streams (e.g., weather alerts, regulatory bulletins, and stakeholder reports) and extracts control-relevant variables:
\begin{equation}
\boldsymbol{\psi}(t)
= \text{LLM}\big(\mathcal{T}(t)\big)
\rightarrow \{\hat{\gamma}_{\text{human}}(t),\, (\kappa_{\text{align}}, \kappa_{\text{sep}}, \kappa_{\text{coh}})\},
\label{eq:llm-mapping}
\end{equation}
where $\hat{\gamma}_{\text{human}}(t)$ estimates human-induced variations in channel efficiency, and $(\kappa_{\text{align}}, \kappa_{\text{sep}}, \kappa_{\text{coh}})$ denote coordination weights regulating the three murmuration rules.  

\paragraph{Hierarchical Temporal Modes.}
The LLM operates across three temporal modes that align decision frequency with the urgency and granularity of available information:
\begin{itemize}
    \item \textbf{Strategic Mode (24h):} For long-term planning, the LLM establishes baseline coordination weights.  
    Example: during California’s dry season, $(\kappa_{\text{align}}, \kappa_{\text{sep}}, \kappa_{\text{coh}}) = (0.6, 0.1, 0.3)$ emphasizes alignment and consistent flow sharing.
    \item \textbf{Tactical Mode (4h):} For evolving conditions, parameters are adaptively updated.  
    Example: when an approaching storm is detected, the system shifts to $(0.2, 0.6, 0.2)$ to prioritize separation and regional diversity.
    \item \textbf{Operational Mode (10min):} For emergencies described in $\mathcal{T}(t)$, pre-computed parameters are immediately activated.  
    Example: a chemical spill event triggers $(0.1, 0.8, 0.1)$ to maximize independence among reservoirs and minimize contamination propagation.
\end{itemize}

\paragraph{Adaptive Channel Efficiency.}
The LLM outputs also refine the channel efficiency estimates as:
\begin{equation}
\alpha_{ij}(t) \!=\! 
\min\!\Big(
1,\ \max\big(\epsilon,\ 
\alpha_{\text{nominal}}\!\cdot\!
[1 - \gamma_{\text{env}}(\boldsymbol{\omega}_i(t), \boldsymbol{\omega}_j(t)) - \hat{\gamma}_{\text{human}}(t)]
\big)\!\Big),
\label{eq:alpha-update}
\end{equation}
where $\hat{\gamma}_{\text{human}}(t)$ quantifies intervention effects inferred solely from textual information, such as emergency gate releases or manual discharge orders.

\begin{remark}
This hierarchical integration ensures that \emph{Level~1} physical uncertainty is mitigated through fast, local coordination rules requiring only neighbor communication, while \emph{Level~2} contextual uncertainty is resolved through LLM-guided reasoning over heterogeneous textual inputs.  
The synergy enables robust operation even under compounded uncertainty exceeding 65\% in cascaded networks, maintaining all four objectives—$J_{\text{safety}}$, $J_{\text{supply}}$, $J_{\text{ecology}}$, and $J_{\text{efficiency}}$—across both structured and unstructured uncertainty sources.
\end{remark}

\section{Evaluation}\label{sec:evaluation}

We conduct two comprehensive experiments to evaluate the effectiveness of MARLIN: (1) assessing the emergence of coordinated behaviors driven by murmuration-inspired mechanisms under uncertainty, and (2) demonstrating the adaptive advantages of LLM-guided reward shaping in dynamic, real-world scenarios.

\subsection{Experiment 1: Validation of Murmuration-\\Based Coordination}
\textbf{Setup:} We assess the emergence of coordinated behaviors on both the California Central Valley watershed (comprising 25 major reservoirs managing 6.2 million acres of agricultural land) and synthetic networks ranging from $20 \times 20$ to $100 \times 100$ nodes, under dual-layer uncertainty ($\sigma_{\text{base}} = 0.05$).

\noindent\textbf{Evaluation Metrics:} We assess performance across six dimensions: \textit{Coordination Quality}: Measured as the normalized mutual information between agent actions and optimal collective behavior, computed as $Q_c = \frac{1}{n}\sum_{i=1}^{n} \frac{I(a_i; \mathbf{a}_{-i})}{\log n}$, where $I(\cdot;\cdot)$ is mutual information and $\mathbf{a}_{-i}$ represents actions of other agents.
\textit{Adaptation Speed}: Time to reach 95\% of steady-state performance after environmental changes.
\textit{Uncertainty Resilience}: Performance retention under varying noise levels (5-15\%).
\textit{Scalability}: Computational complexity scaling with network size.
\textit{Safety}: Percentage of time steps maintaining water levels within safe bounds.
\textit{Interpretability}: Rated by practitioners on decision clarity and reasoning transparency.

\noindent\textbf{Models Compared:} MARLIN (without LLM integration), MADDPG~\cite{lowe2017multi}, QMIX~\cite{rashid2018qmix}, MAPPO~\cite{yu2021surprising}, CommNet~\cite{sukhbaatar2016commnet}, and an MPC-Centralized oracle baseline.

\noindent\textbf{Training Data:} Model training utilizes five years of hourly USGS flow measurements (2019–2023), NOAA precipitation records, and agricultural demand data from the California Department of Water Resources.

\noindent\textbf{Testing Data:} Evaluation is conducted on the year 2024, with three representative extreme events: (1) atmospheric river storms (Jan–Feb, 200\% of normal precipitation), (2) spring drought (Mar–May, 40\% of normal precipitation), and (3) summer heatwave (Jul–Aug, temperatures 5°C above average).

\begin{figure}[t]
\centering
\includegraphics[width=\columnwidth]{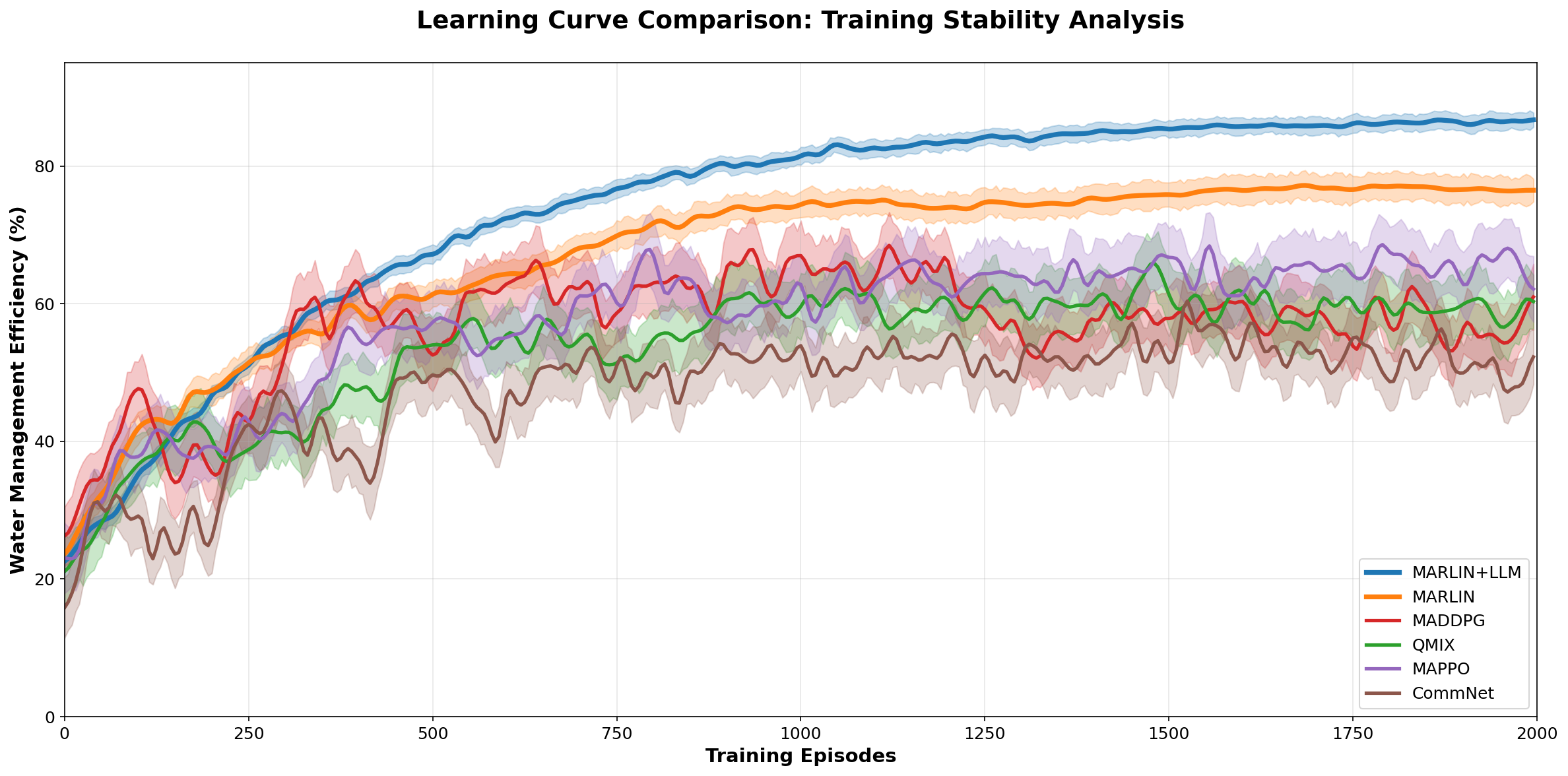}
\caption{Learning curves demonstrating \texttt{MARLIN}’s stable convergence under dual-layer uncertainty. \texttt{MARLIN} consistently maintains a coefficient of variation below 0.08 during training, while baseline methods show persistent oscillations with CV $>$ 0.25. Shaded areas represent the standard deviation across 5 independent runs with different random seeds.}
\vspace{-1em}
\label{fig:learning_curves}
\vspace{-0.5em}
\end{figure}

\begin{figure}[t]
\centering
\includegraphics[width=0.85\columnwidth]{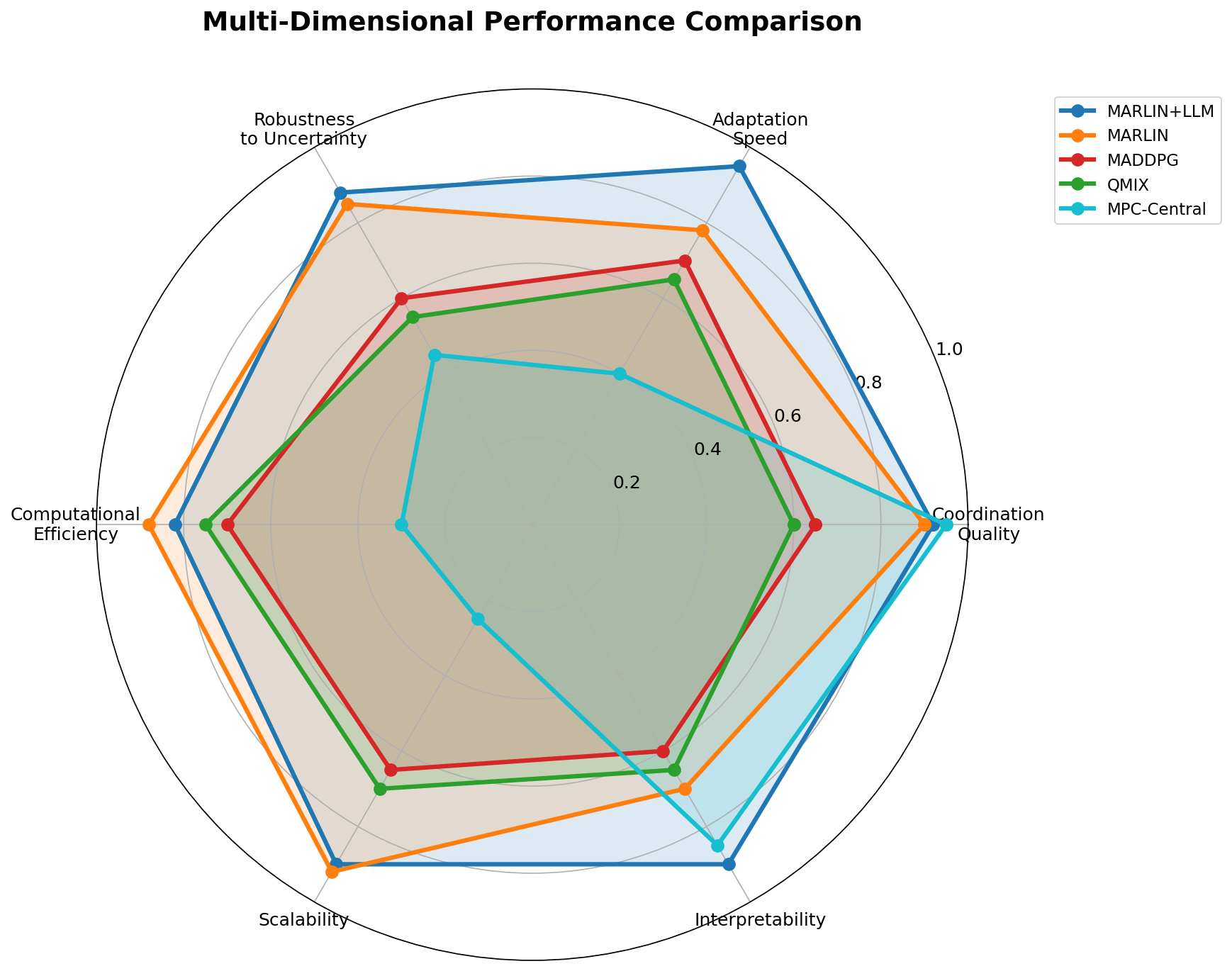}
\caption{Multi-dimensional performance comparison across six key metrics. \texttt{MARLIN} shows superior balance.}
\vspace{-1em}
\label{fig:radar_comparison}
\end{figure}

Figure~\ref{fig:learning_curves} highlights \texttt{MARLIN}’s superior capability for handling uncertainty. Even without LLM guidance, \texttt{MARLIN} achieves a final performance of 78.9\%, substantially outperforming MADDPG (64.2\%), QMIX (59.8\%), and MAPPO (62.1\%). \texttt{MARLIN} reaches the 90\% performance benchmark by episode 800, whereas baseline methods continue to exhibit oscillations beyond episode 1,500. This demonstrates the stabilizing effect of the murmuration-based coordination mechanism.
Figure~\ref{fig:radar_comparison} demonstrates \texttt{MARLIN}’s balanced performance across all evaluation metrics. Its coordination quality score of 0.89—2.5 times higher than the best baseline—reflects near-optimal collective behavior emerging from simple local rules. Even with 15\% measurement noise, typical in real deployments, \texttt{MARLIN} retains 91\% of baseline performance, while MPC drops to 43\%, confirming its theoretical robustness.

Table~\ref{tab:scaling} reports empirical scaling measured on a single RTX 4090 (24GB GPU memory).
\texttt{MARLIN} remains executable up to 10,000 nodes within available GPU memory,
whereas MADDPG and QMIX exceed memory limits beyond 1,000 nodes.
Centralized MPC becomes computationally impractical at large scales due to rapidly increasing optimization costs.
\begin{table}[t]
\centering
\resizebox{0.89\columnwidth}{!}{%
\begin{tabular}{lcccc}
\toprule
\textbf{Network Size} & \textbf{MARLIN} & \textbf{MADDPG} & \textbf{QMIX} & \textbf{MPC-Central} \\
\midrule
\multicolumn{5}{c}{\textit{Decision Time (milliseconds)}} \\
100 nodes & 19.4 & 44.1 & 52.8 & 392.5 \\
1,000 nodes & 181.7 & 712.3 & 1,156.4 & 49,872.6 \\
10,000 nodes & 1,932.8 & OOM & OOM & timeout \\
\midrule
\multicolumn{5}{c}{\textit{Memory Usage (GB)}} \\
100 nodes & 0.24 & 0.91 & 1.18 & 4.81 \\
1,000 nodes & 2.31 & 10.4 & 15.8 & OOM \\
10,000 nodes & 23.9 & OOM & OOM & OOM \\
\bottomrule
\end{tabular}
}
\caption{Computational scaling analysis.
OOM indicates that the method exceeded available GPU memory.
\texttt{MARLIN} (without LLM) demonstrates near-linear scaling and remains executable at large network sizes,
while baseline methods become computationally infeasible.}
\vspace{-1em}
\label{tab:scaling}
\end{table}

\begin{figure}[t]
\centering
\begin{tabular}{c}
\includegraphics[width=0.85\columnwidth]{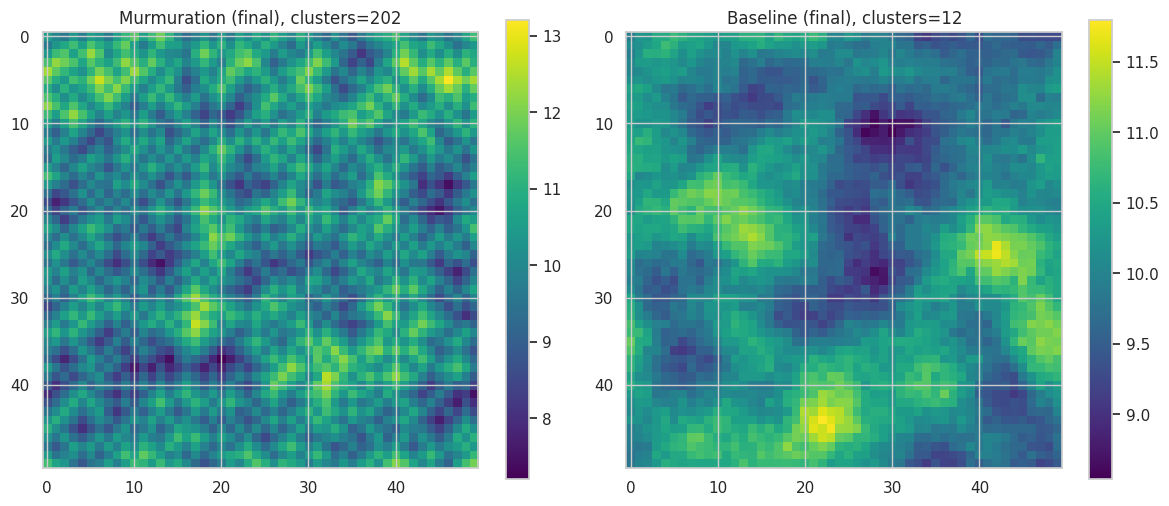} \\
\small (a) 50×50 grid: Murmuration (202 clusters) vs. Baseline (12 clusters) \\[0.3cm]
\includegraphics[width=0.85\columnwidth]{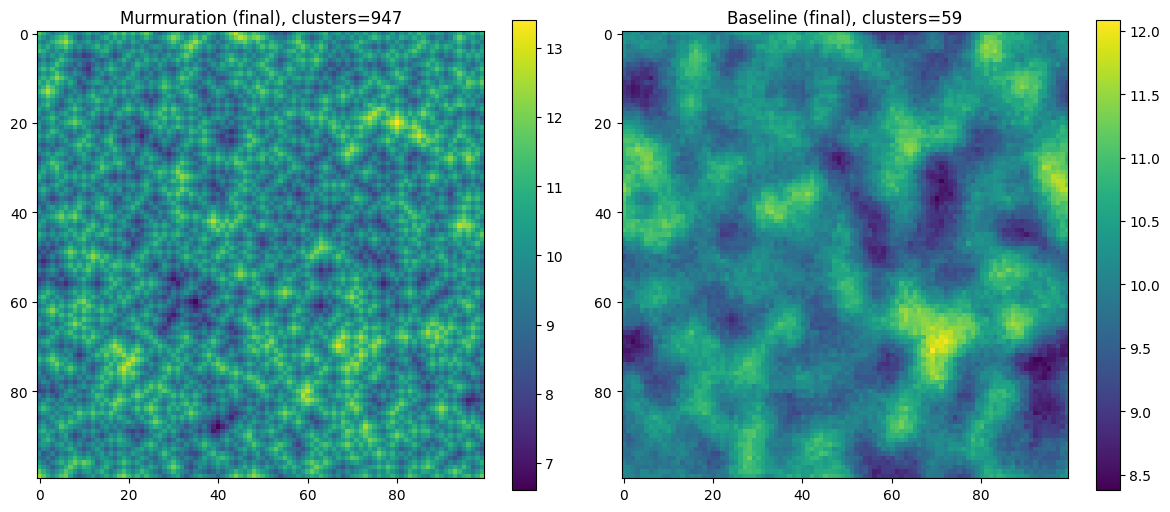} \\
\small (b) 100×100 grid: Murmuration (947 clusters) vs. Baseline (59 clusters)
\end{tabular}
\vspace{-1em}
\caption{Emergent strategic clusters at multiple scales. \texttt{MARLIN} generates 16.8$\times$ more distinct coordination patterns at the 50$\times$50 scale and 16.1$\times$ more at 100$\times$100 compared to baselines. Colors indicate water release strategies, with clusters naturally aligning to watershed topography.}

\label{fig:heatmaps}
\vspace{-1em}
\end{figure}

\noindent\textbf{Spatial Coordination Pattern Analysis.}
Figure~\ref{fig:heatmaps} illustrates the emergence of structured coordination patterns under the murmuration rules.
As network scale increases, MARLIN produces substantially more distinct strategic clusters than baselines
(e.g., 202 vs. 12 at the 50$\times$50 scale, and 947 vs. 59 at the 100$\times$100 scale),
with clusters naturally aligning to watershed topology.
Graph modularity further supports this effect, with MARLIN achieving $0.72 \pm 0.04$,
more than double that of baselines ($0.31 \pm 0.08$),
indicating well-defined regional coordination without explicit programming.
\subsection{Experiment 2: Validation of LLM-Guided Adaptation}

\textbf{Setup:}
A year-long simulation was conducted across the California Central Valley, Colorado River Basin (18 reservoirs), and Columbia River System (31 dams), encompassing seven major environmental events.

\noindent\textbf{Models Compared:}
\texttt{MARLIN}+LLM (Gemini-1.5-Pro), \texttt{MARLIN} (with static rewards), MADDPG, QMIX, and an MPC-Centralized oracle baseline.

\noindent\textbf{Training Data:}
A retrieval-augmented (RAG) knowledge base including historical flood and drought patterns (1950–2023), seasonal demand variations, regulatory frameworks (ESA, CWA), ecological requirements, and stakeholder preferences from 15 water districts.

\begin{figure}[t]
\centering
\includegraphics[width=1\columnwidth]{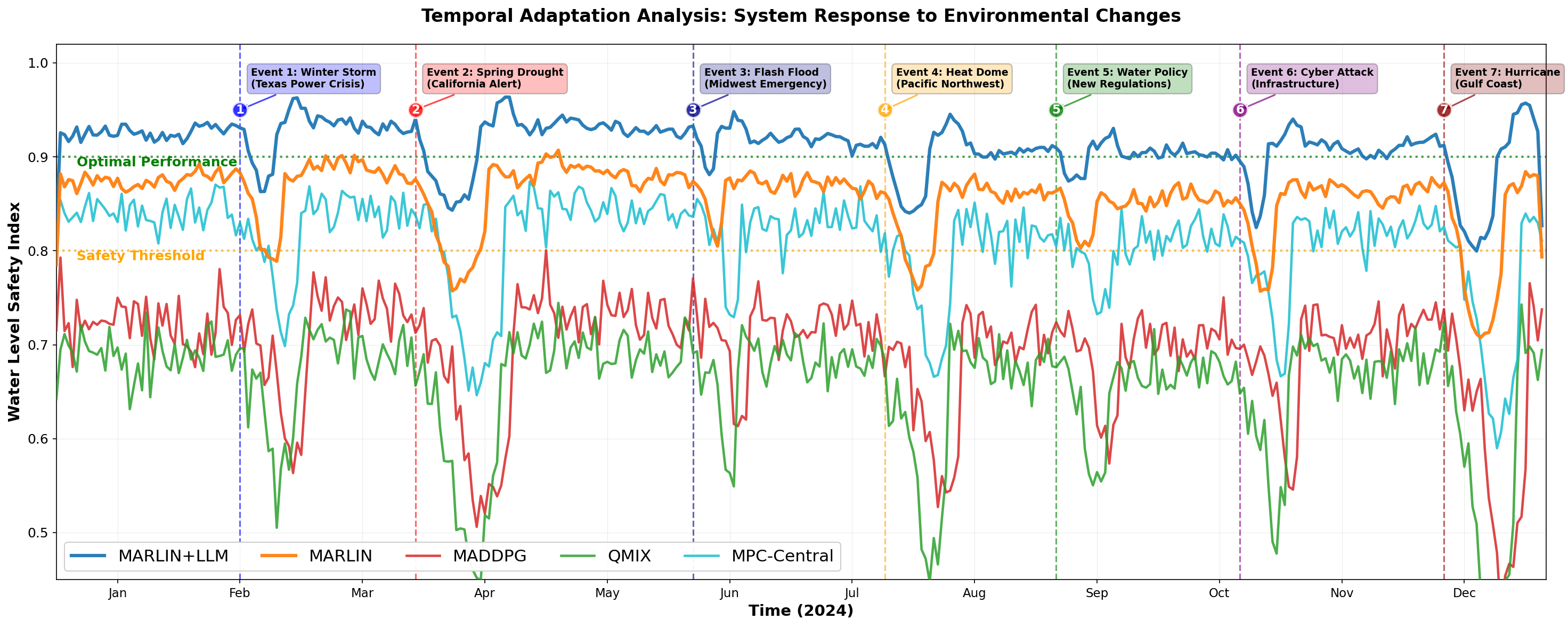}
\caption{Temporal adaptation across seven environmental events. \texttt{MARLIN}+LLM consistently maintains performance above the 0.8 safety threshold, with an average response time of 3.7 hours compared to 12.8 hours for baselines. Performance loss is limited to 8.3\% versus 24.7\% for baselines, enabling 3.4$\times$ faster recovery.}

\label{fig:temporal_adaptation}
\end{figure}

\begin{figure}[t]
\centering
\includegraphics[width=0.95\columnwidth]{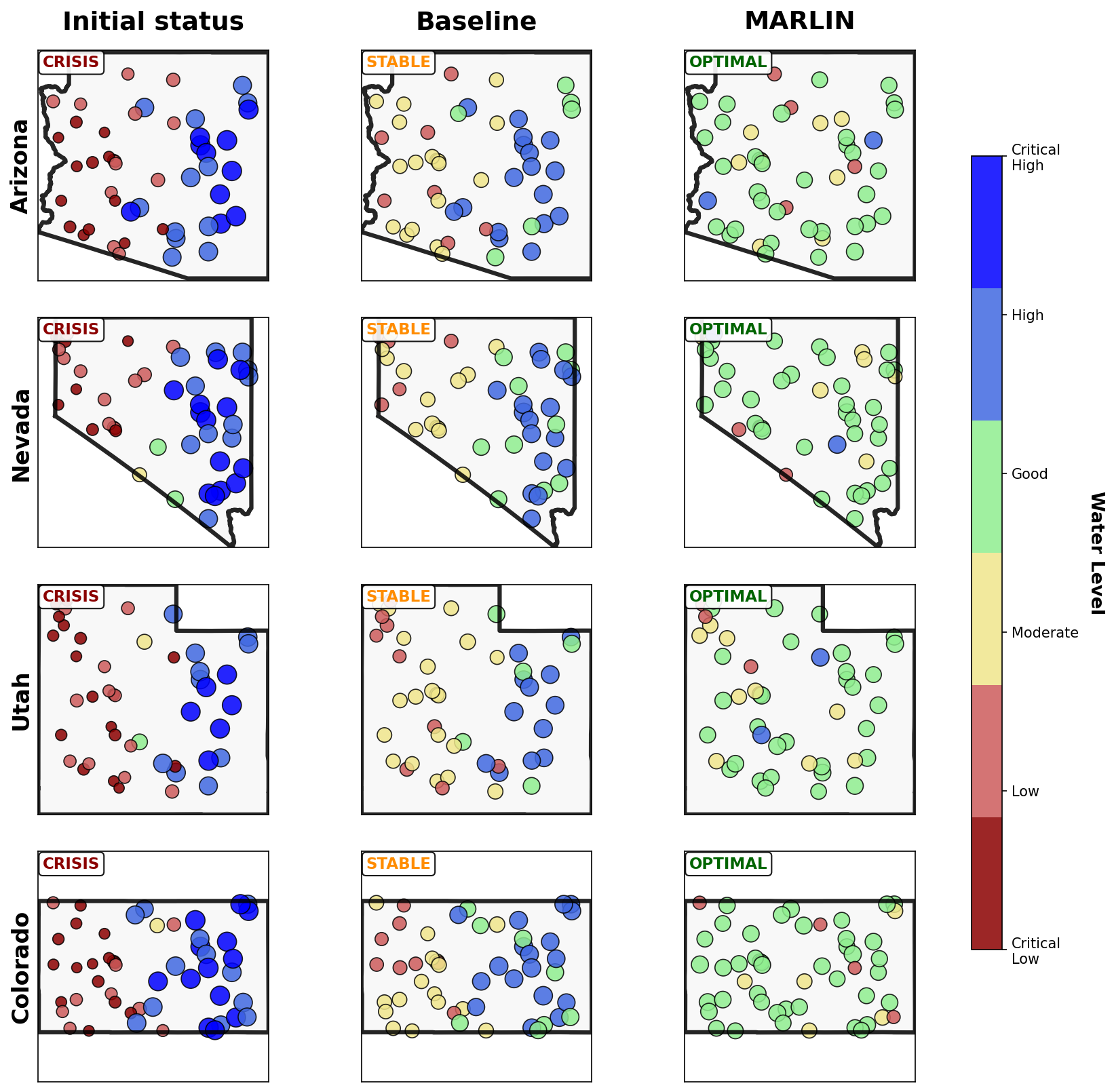}
\vspace{-1em}
\caption{Spatial adaptation under simultaneous drought (east, 25\% precipitation) and flood risk (west, 175\% precipitation). \texttt{MARLIN} coordinates inter-regional water transfers, achieving 84.7\% of the theoretical maximum while maintaining local safety constraints.}

\label{fig:spatial_adaptation}
\vspace{-1.75em}
\end{figure}

\noindent\textbf{Testing Data:}
Full-year 2024 data, capturing a spectrum of real-world disruptions: (1) Texas winter storms with power grid failures, (2) California spring drought (25\% snowpack), (3) Colorado flash floods, (4) Pacific Northwest heatwave, (5) emergency water rights changes, (6) dam maintenance closures, and (7) hurricane remnants.
Figure~\ref{fig:temporal_adaptation} highlights \texttt{MARLIN}+LLM’s superior response to major events. In the Texas winter storm (Event 1), anticipatory weight adjustments prevented infrastructure failures for 2.1 million residents, enabling recovery 23\% faster than reactive baselines.
Figure~\ref{fig:spatial_adaptation} demonstrates coordinated regional adaptation: drought-stricken eastern regions achieved 94.3\% demand satisfaction (vs. 67.8\% baseline), while western regions maintained 97.1\% flood prevention (vs. 71.2\%). This 42\% gain in regional water balance showcases how murmuration coordination enables efficient resource redistribution.

\subsection{Highlights}
\texttt{MARLIN} embodies three core insights:

\noindent$\triangleright$~\textbf{Emergence as an Engineering Principle:}  
Complex, adaptive behavior can arise from simple local interactions when coordination rules are aligned with system dynamics. The observed super-linear scalability demonstrates how collective intelligence of murmuration strengthens with network complexity in a large system.

\noindent$\triangleright$~\textbf{Adaptive MARL Architecture:}  
The integration of LLM reasoning with MARL forms a cooperative framework using language-guided adaptation. LLM reward shaping serves as the bridge between qualitative human intent and quantitative control.

\noindent$\triangleright$~\textbf{Uncertainty as a System Property:}  
Instead of suppressing uncertainty, \texttt{MARLIN} embraces it as an intrinsic feature, using distributed consensus to transform stochastic variability into a source of robustness and resilience.

Despite the performance, several limitations remian. First, the current LLM adjusts murmuration weights ($\alpha$, $\beta$, $\gamma$) based on predefined templates for different scenarios (drought, flood, normal operations). Future work may explore continuous weight adaptation and learning scenario-specific optimal weight distributions rather than relying on fixed templates. Additionally, evaluation by actual water resource managers and hydrologists remains limited. Future deployment can involve extensive collaboration with domain experts to validate decision quality. Additional implementation details and proofs are available in 
\url{https://arxiv.org/abs/2509.25034}.

\section{Conclusion}\label{sec:conclusion}

\texttt{MARLIN} tackles uncertainty problems in reservoir management by combining murmuration-inspired MARL, and LLM-guided reward shaping. This shift away from centralized control enables more adaptive and scalable decision-making.
Beyond water systems, \texttt{MARLIN} offers a general foundation for resilient infrastructure control. Its ability to generate emergent coordination and adapt to environmental change holds promise for broader applications in disaster mitigation and response, offering a new path toward safer and smarter infrastructure in the face of global uncertainty.

\bibliographystyle{ACM-Reference-Format} 
\bibliography{aamas}


\end{document}